\documentclass[1p,times,procedia]{elsarticle}
\usepackage{nupha_ecrc}

\volume{00}

\firstpage{1}

\journalname{Nuclear Physics A}

\runauth{C. Shen el al.}

\jid{nupha}

\jnltitlelogo{Nuclear Physics A}

\usepackage{amssymb}

\usepackage[figuresright]{rotating}

\begin{document}

\begin{frontmatter}



\dochead{The XXVth International Conference on Ultrarelativistic Nucleus-Nucleus Collisions}

\title{Direct photon production and jet energy-loss in small systems}


\author[1]{Chun Shen}
\author[1]{Chanwook Park}
\author[1,2]{Jean-Fran\c{c}ois Paquet}
\author[1,3]{Gabriel S. Denicol}
\author[1]{Sangyong Jeon}
\author[1]{Charles Gale\footnote{Speaker.}}

\address[1]{Department of Physics, McGill University, 3600 University Street, Montreal, QC, H3A 2T8, Canada}
\address[2]{Department of Physics \& Astronomy, Stony Brook University, Stony Brook, NY 11733, USA}
\address[3]{Physics Department, Brookhaven National Laboratory, Upton, NY 11973, USA}

\begin{abstract}
Two types of penetrating probes, direct photon and QCD jets, are investigated in the background of a small and rapidly expanding droplet of quark-gluon plasma. The additional thermal electromagnetic radiation results in a $\sim$50\% enhancement of the direct photons. In  high multiplicity p+Pb collisions, jets can lose a sizeable fraction of their initial energy, leading to a charged hadron $R_\mathrm{pA}$ of $\sim$0.8 at a transverse momentum around 10\,GeV. Those two proposed measurements can help understand the apparent collective behaviour observed in small collision systems.
\end{abstract}


\end{frontmatter}


\section{Introduction}
\label{intro}

Signatures usually associated with hydrodynamic behaviour have been recently observed in high and intermediate multiplicity proton-nucleus (p+Pb) collisions at the Large Hadron Collider\,\cite{ABELEV:2013wsa,Khachatryan:2015waa,Aad:2014lta}. Even though those measurements are suggestive of the creation of a strongly coupled quark-gluon plasma (QGP), additional confirmation involving complementary observables needs to be pursued. 

In this contribution we calculate the thermal photon radiation produced by a small and rapidly expanding QGP droplet, and we evaluate the energy loss of high energy QCD jets. We find that a significant amount of  electromagnetic radiation is emitted, with thermal photons accounting for $\sim$50\% of the direct photons produced in high multiplicity p+Pb collisions at low $p_T$. Furthermore, we show that in spite of the small system size, jets still lose a sizeable fraction of their initial energy, leading to a charged hadron $R_\mathrm{AA}$ of 0.8 at a transverse momentum of $\sim$10\,GeV. Should these two measurements be within  reach of the current generation of experiments, they would serve as additional evidence that a strongly coupled QGP is being produced in proton-nucleus collisions at the LHC. To complete the analysis, we also study direct photon production in other small systems, such as (p, d $^3$He)+Au collisions at  top RHIC energy.

\section{Direct photons} 

\begin{figure*}[h!]
\centering
\begin{tabular}{ccc}
  \includegraphics[width=0.30\linewidth]{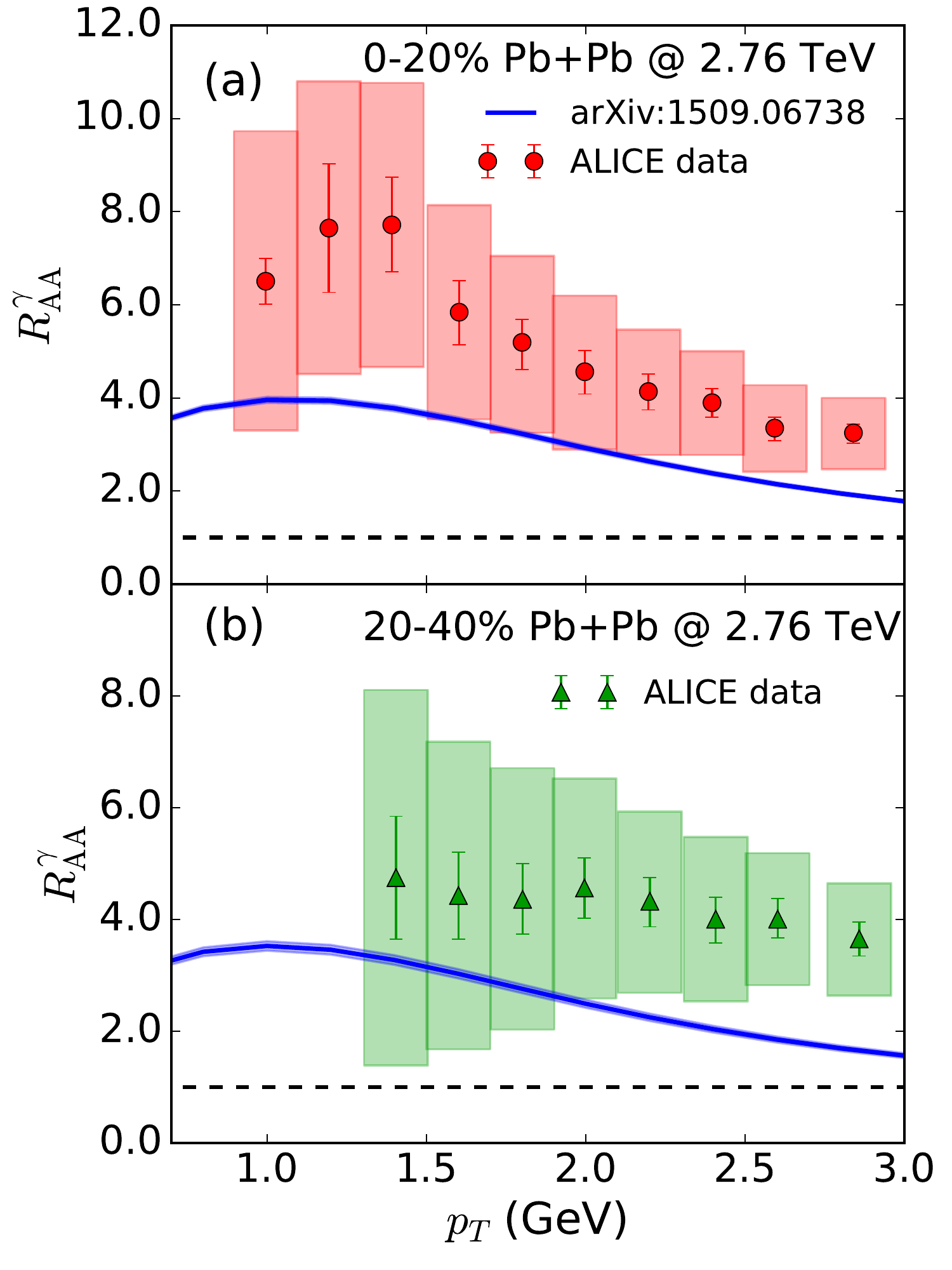} & 
  \includegraphics[width=0.30\linewidth]{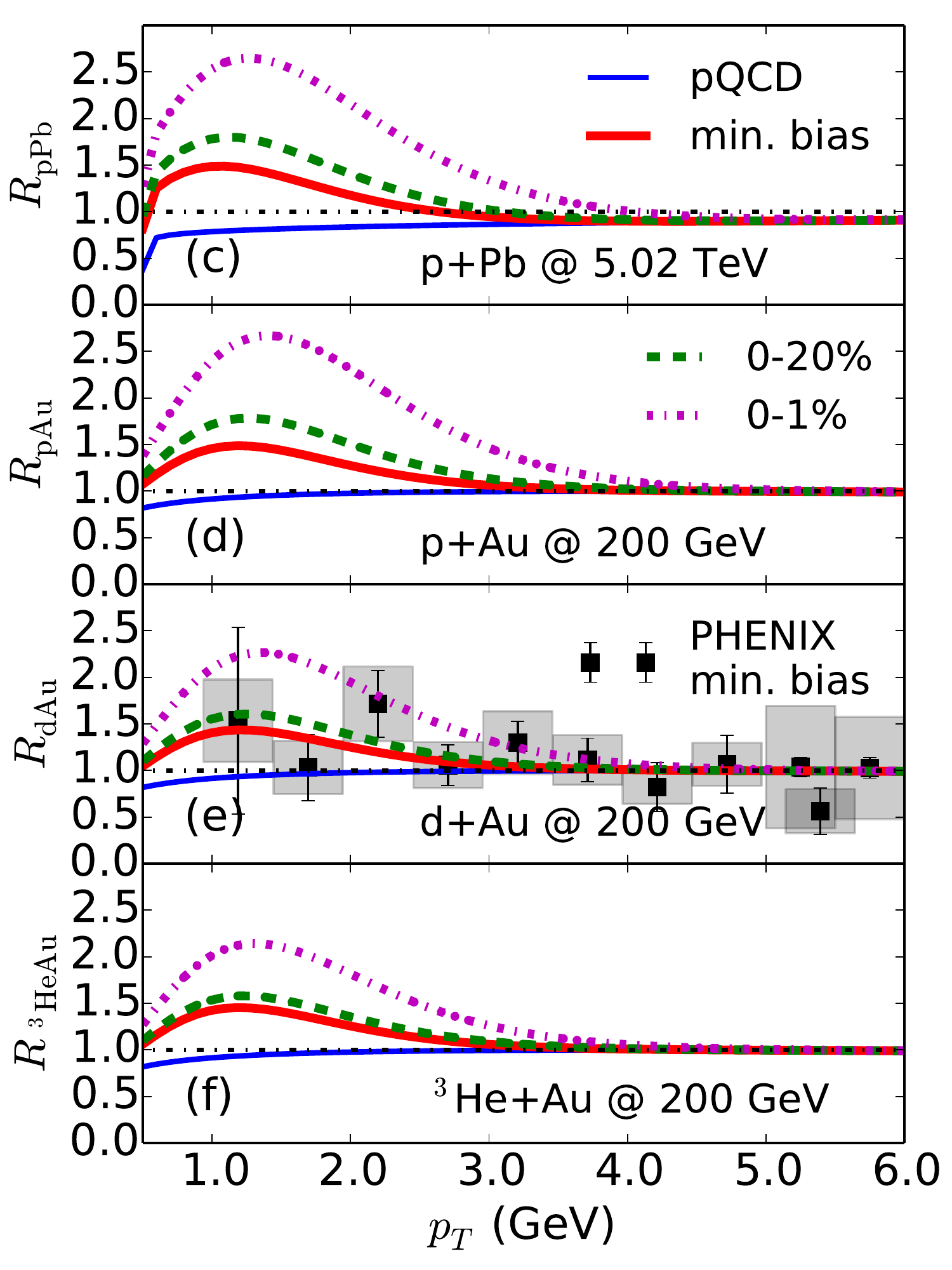} & 
  \includegraphics[width=0.30\linewidth]{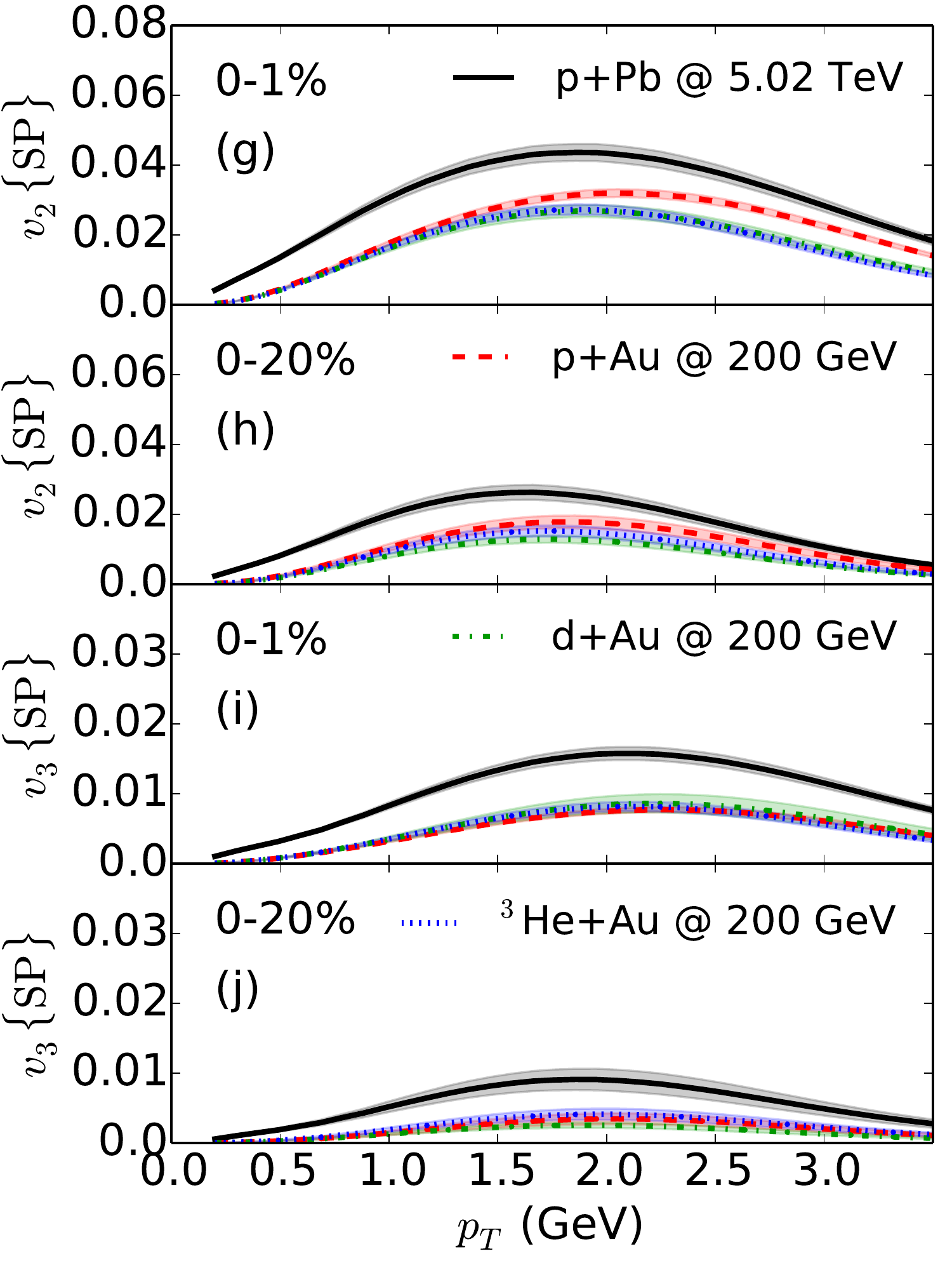}
\end{tabular}
\caption{{\it Panel (a)-(b):} Direct photon $R^\gamma_\mathrm{AA}$ in Pb+Pb collisions at LHC energy. {\it Panel (c)-(f):} The nuclear modification factor of direct photons in central and minimum bias collisions in small systems. {\it Panel (g)-(j):} The anisotropic flow coefficients $v_{2,3}\mathrm{SP}$(scalar-product method) for direct photons in 0-1\% and 0-20\% centrality bins. }
\label{fig1}
\end{figure*}

    In Pb+Pb collisions at $\sqrt{s_\mathrm{NN}} = 2.76$\,TeV, the measured direct photon yield shows a large enhancement over the  photon spectra measured in pp collisions  scaled by $N_\mathrm{coll}$, the number of binary collisions  \cite{Adam:2015lda,Sahlmueller:2015rhy}. This is usually quantified by the nuclear modification factor $R_\mathrm{AA}(p_T) \equiv \frac{dN^\mathrm{AA}}{dy dp_T}/(N_\mathrm{coll} \frac{dN^\mathrm{pp}}{dy dp_T})$. Direct photon $R_\mathrm{AA}$ in Pb+Pb collisions are shown in Figs.\,\ref{fig1}a-b, where it is seen that most of this enhancement - if not all - can be attributed to thermal radiation from the bulk medium \cite{Paquet:2015lta}. Similarly, if a smaller medium achieves near-equilibrium, it will radiate thermal photons during its evolution. We investigated the thermal photon radiation from p+Pb collisions at $\sqrt{s_\mathrm{NN}} = 5.02$\,TeV and (p, d, $^3$He)+Au at $\sqrt{s_\mathrm{NN}} = 200$\,GeV using the publicly available {\tt iEBE-VISHNU} framework\,\cite{Shen:2014vra,Shen:2015qba}. Event-by-event hydrodynamic simulations  with fluctuating Monte-Carlo Glauber initial conditions were used. Additional collision-by-collision fluctuations were considered in the initial energy density profiles, which led to a better description of charge hadron multiplicity distributions for these small collision systems\,\cite{Shen:2014vra}. The collision centrality was determined by the initial total entropy, $dS/dy$, which serves as a good approximation of the charged hadron multiplicity\,\cite{Shen:2015qta}. 
%
%

In Figs.\,\ref{fig1}c-f, we consider the direct photon signals below $p_T < 3$\,GeV, and find that  the thermal enhancement was about 50\% in minimum bias collisions, and increased up to $\sim$150\% for collisions in the 0-1\% centrality class. Even though the space-time volumes of the proton-induced collisions are 20 times smaller than those in Pb+Pb collisions, the thermal signal is only smaller by about a factor of 2. The  direct photon measurements in minimum bias d+Au collisions still contain substantial statistical and systematic uncertainties\cite{Adare:2012vn}. Currently, none of the options - with or without thermal enhancement - are excluded by the data. The direct photon $R^\gamma_{AA}$ increase as a function of $p_T$ is faster for central collisions in p+Pb at $\sqrt{s_\mathrm{NN}} = 5.02$\,TeV and p+Au at $\sqrt{s_\mathrm{NN}} = 200$\,GeV than in (d, $^3$He)+Au collisions at the top RHIC energy, owing to the fact that the number of fluctuating sources is smaller for p+A than for (d, $^3$He)+A. Event-by-event fluctuations are more pronounced and will lead to higher temperatures in central p+A collisions. By correlating direct photon with charged hadrons, the direct photon anisotropic flow coefficients, $v_{2,3}\{\mathrm{SP}\}$, in 0-1\% and 0-20\% centrality bins are shown in Fig.\,\ref{fig1}g-j. The direct photon anisotropic flow in p+Pb collisions at 5.02 TeV is the largest among the four  systems considered, because of larger pressure gradients and longer fireball lifetime.

\section{Jet energy loss}

Unlike direct photons that penetrate the fireball unscathed, high-energy partonic jets are strongly interacting probes. They lose some of their energy  passing through the QGP. 
We estimate the amount of jet energy-loss in p+Pb collisions at $\sqrt{s_\mathrm{NN}} = 5.02$\,TeV using the Monte-Carlo event generator {\tt MARTINI}\,\cite{Schenke:2009gb}. Initial jets were generated using {\tt Pythia}\,\cite{Sjostrand:2014zea} with the standard tune settings\,\cite{Skands:2014pea}. Partons inside the proton were sampled from the parton distribution function (PDF), CTEQ6\,\cite{Nadolsky:2008zw}. For the Pb nucleus, additional cold nuclear effects were included by using EPS09 nuclear PDF\,\cite{Eskola:2009uj}. These energetic partons were then evolved against the event-by-event hydrodynamic medium. Medium-induced radiation, elastic collisions, and path length dependence effects were taken into account in the model simulations\,\cite{Schenke:2009gb,Jeon:2003gi,CaronHuot:2010bp}. Because the medium temperature is lower than perturbative scale, we choose a fixed strong coupling constant $\alpha_s = 0.27$ between jets and medium interaction. The strong coupling constant associated with medium-induced radiation is set to be running according to the jet momentum. All the partons were evolved to $T_c = 165$\,MeV and then hadronized according to the string fragmentation scheme germane to {\tt Pythia}. 

%
\begin{figure*}[h!]
\centering
\begin{tabular}{cc}
  \includegraphics[width=0.43\linewidth]{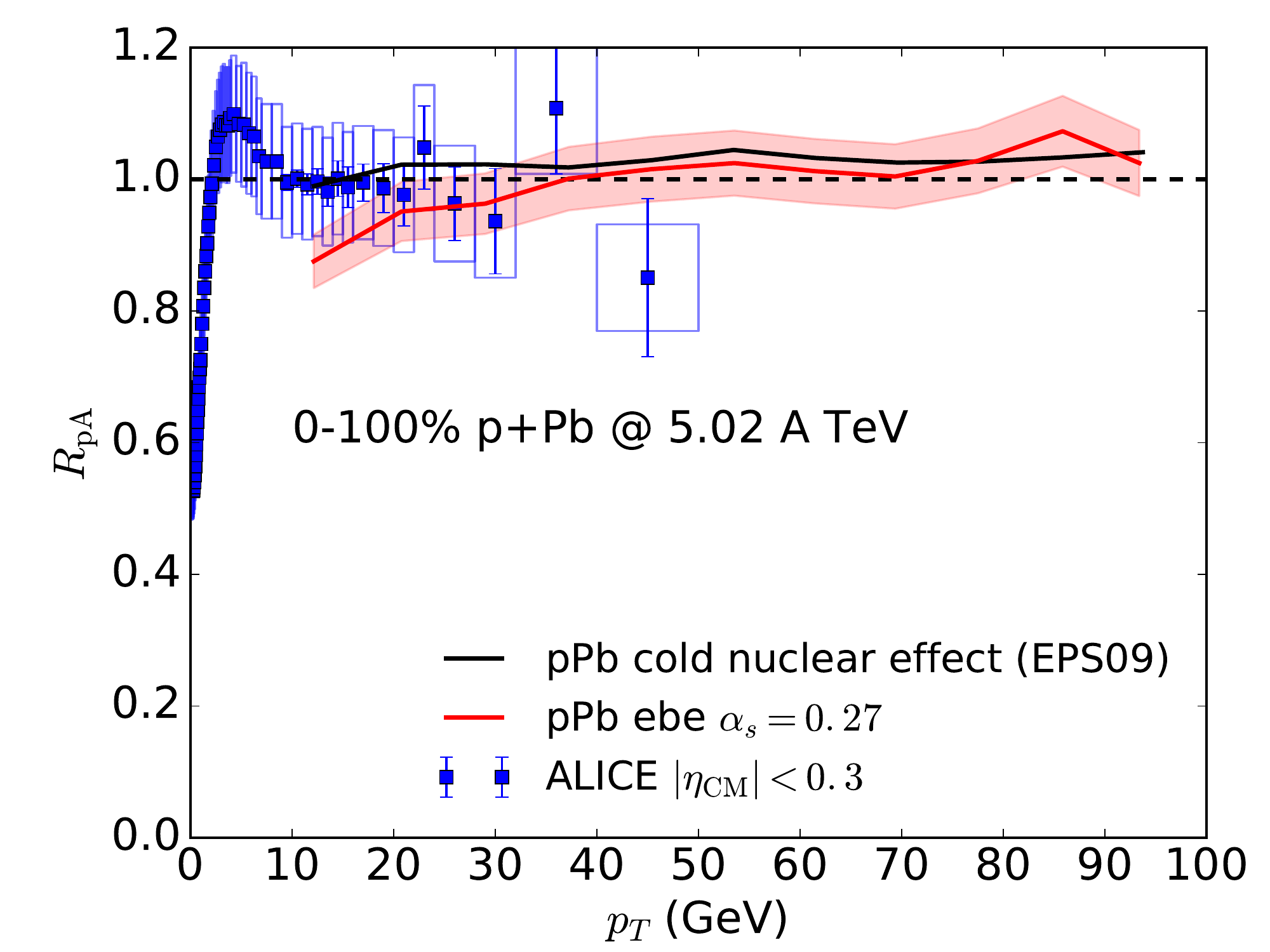} & 
  \includegraphics[width=0.43\linewidth]{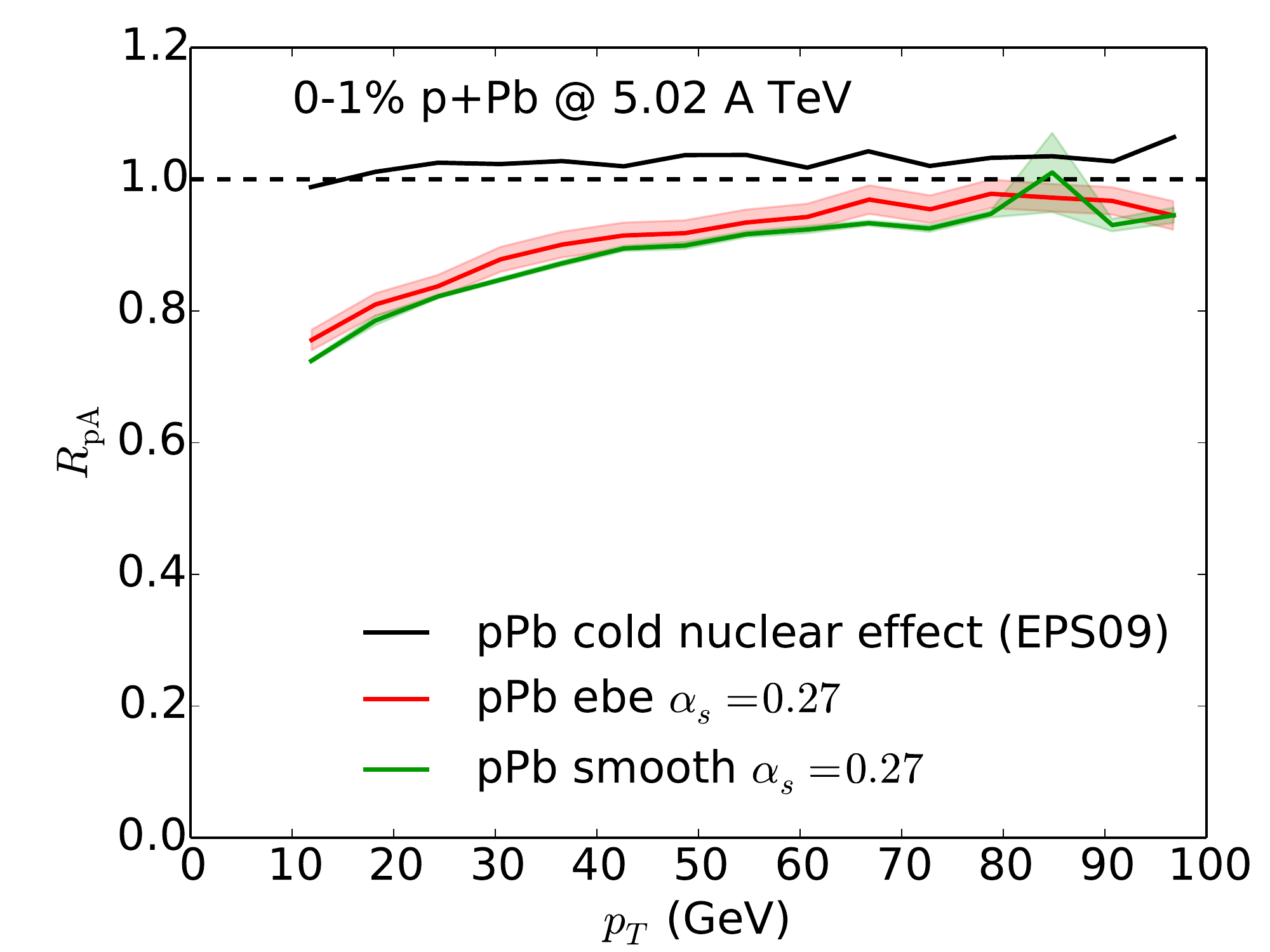}
\end{tabular}
\caption{The nuclear modification factor of charged hadron, $R_\mathrm{pA}$, in minimum bias and 0-1\% centrality p+Pb collisions at $\sqrt{s_\mathrm{NN}} = 5.02$\,TeV. Results in minimum bias collisions are compared with the ALICE measurement\,\cite{ALICE:2012mj}.}
\label{fig2}
\end{figure*}
%

In Fig.\,\ref{fig2}, charged hadron $R_\mathrm{pA}$ in minimum bias and 0-1\% centrality p+Pb collisions are shown for $10 < p_T < 100$\,GeV. In minimum bias collisions, the effect of energy loss is found to be negligible for $p_T > 40$\,GeV; only a modest mount of suppression is observed for $10 < p_T < 40$\,GeV. Both  calculations - with and without energy loss - are consistent with the current ALICE data \cite{ALICE:2012mj}. On the other hand, a sizeable jet energy loss  in the 0-1\% central collisions is shown in the right panel of Fig.\,\ref{fig2}. Charged hadron $R_\mathrm{pA}$ is suppressed to $\sim$0.8 for $p_T \sim 10-20$\,GeV, and  increases to $\sim$ 1 around $p_T = 100$\,GeV. We also find that energy loss calculations with a smooth event-averaged hydrodynamic medium yield results for charged hadron $R_\mathrm{pA}$ very similar to those obtained with more realistic event-by-event simulations. The effects of event-by-event fluctuations on charged hadron $R_\mathrm{pA}$ is limited, for p + Pb collisions at the LHC. This is because the size of the entire fireball in the event-averaged hydrodynamic simulation is comparable to the typical total size of the hot spots in a fluctuating medium for p+Pb collisions. The charged hadron $R_\mathrm{pA}$ does not have the resolution to make such a distinction. 

\section*{Acknowledgement}

This work was supported by the Natural Sciences and Engineering Research Council of Canada.  Computations were made on the Guillimin supercomputer at McGill University, managed by Calcul Qu\'{e}bec and Compute Canada. 





\bibliographystyle{elsarticle-num}
\bibliography{references}







\end{document}